# Ransomware Detection using Process Memory

**Avinash Singh[1], Richard Adeyemi Ikuesan[2] and Hein Venter[1]**
**[1]University of Pretoria, South Africa**
**[2]Community College Qatar, Doha, Qatar**
asingh@cs.up.ac.za
richard.ikuesan@ccq.edu.qa
hventer@cs.up.ac.za

**Abstract:** Ransomware attacks have increased significantly in recent years, causing great destruction and damage to critical systems and business operations. Attackers are unfailingly finding innovative ways to bypass detection mechanisms, which encouraged the adoption of artificial intelligence. However, most research summarizes the general features of AI and induces many false positives, as the behavior of ransomware constantly differs to bypass detection. Focusing on the key indicating features of ransomware becomes vital as this guides the investigator to the inner workings and main function of ransomware itself. By utilizing access privileges in process memory, the main function of the ransomware can be detected more easily and accurately. Furthermore, new signatures and fingerprints of ransomware families can be identified to classify novel ransomware attacks correctly. The current research used the process memory access privileges of the different memory regions of the behavior of an executable to quickly determine its intent before serious harm can occur. To achieve this aim, several well-known machine learning algorithms were explored with an accuracy range of 81.38% – 96.28%. The study thus confirms the feasibility of utilizing process memory as a detection mechanism for ransomware.

## 1. Introduction

Detecting ransomware attacks has always been a significant problem for organizations (Singh, Ikuesan & Venter, 2019a,b). Ransomware is the most dominant form of malware that causes major disruptions to daily business operations (Hampton, Baig & Zeadally, 2018; Newman, 2018; Sienko, 2021). The primary objective of ransomware is to encrypt sensitive files and hold the decryption of those files as a ransom for large amounts of money (Singh *et al.*, 2019a). In most instances, ransomware is delivered through major exploits or social engineering. According to the State of Ransomware 2021 report (Sophos, 2021), 54% of organizations that were hit by ransomware had their data encrypted. However, it is estimated that the average ransom payment made by mid-sized organizations was approximately USD 170,404 (Sophos, 2021). The report (Sophos, 2021) also calculated the average cost to recover from a ransomware attack to be around USD 1.8 million. Despite such attacks being a growing concern, the good news is that they have decreased slightly since 2017. This was the time when WannaCry wreaked havoc over the internet with more than 54% of organizations hit, but the number of attacks gradually decreased to around 51% in 2020 and 37% in 2021 (Sophos, 2021).

An alarming fact in the Sophos (2021) report is that 47% of respondents expected ransomware to hit their organization because "Ransomware attacks are increasingly hard to stop due to their sophistication". Even though recovery from ransomware attacks has increased through implicit and immutable backups, there have been significant improvements in methods to detect ransomware attacks. Unfortunately, these methods are easily bypassed by sophisticated ransomware, which leaves security researchers in disarray to find effective detection methods. For instance, to achieve speed and efficiency in detection, anti-virus programs use static and signature-based detection methods. Static analysis is limited to known signatures and snippets of codes. Furthermore, the concept of signature-based detection suggests that only known attributes and unique behaviour can be identified. Thus, both signature-based and static analysis are not efficient against novel ransomware, making these methods inefficient in detecting ransomware attacks. This is due to ransomware being more behavioural based. Since ransomware utilizes security features like encryption, it is difficult to detect new never-seen-before ransomware (Bromium Labs, Kotov & Rajpal, 2014). To solve these issues, researchers have adopted Artificial Intelligence (AI) to detect ransomware attacks more accurately. However, there are many ways AI can be used to detect ransomware, and many researchers make use of automated feature extraction. Certain aspects may sometimes be overlooked, as the features are sourced from techniques and patterns/trends seen in the dataset that the researchers have – thus making it less accurate or slightly biased. It is important for experts or through rigorous experimentation to identify and hand select features that would provide the most insight and that cannot be mocked or influenced. In this paper, process memory was examined to determine how accurately it can detect ransomware. Process memory is behaviour-based and one of the core functions of ransomware to encrypt files (Subedi, Budhathoki & Dasgupta, 2018).

Going forward, Section 2 provides the necessary background and related works, while Section 3 details the proposed method and results of ransomware detection. Section 4 evaluates the performance of the proposed method. The conclusion and future works are presented in Section 5.

## 2. Related Work

Over the years, significant advances have been made in ransomware detection, especially after the devastation that WannaCry caused in 2017 (Adamov & Carlsson, 2017; Berrueta, Morato, Magana, *et al.*, 2020; Fernando, Komninos & Chen, 2020; Molina, Torabi, Sarieddine, *et al.*, 2021; Singh *et al.*, 2019a). Although researchers explored avenues for detection such as static and dynamic information, ransomware has managed to evade static analysis (Subedi *et al.*, 2018). Due to this limitation, typically active behavioural-based detection techniques are used, such as API calls, registry information, opcodes, manual analysis, and dissection (Molina *et al.*, 2021). Manual analysis is quite time-consuming and has led to several sandbox environments with analysis packages like Cuckoo Sandbox (Guarnieri, 2014), which is widely used by security researchers. Machine Learning (ML) has been widely utilized for the detection of malware as well as ransomware (Dang, Di Troia & Stamp, 2021; Dua & Du, 2016; Singh & Singh, 2021). For example, Blackberry Cylance Anti-Virus leverages ML techniques to provide more robust production against unknown attacks (BlackBerry, 2021). This causes challenges with detection rates and the way detection happens, as researchers were able to bypass Cylance protection (SCMagazine, 2019). There is a plethora of research that uses various features for ransomware detection, for example, the work by Ashraf *et al.* (2019) explores static and dynamic information for detection such as API calls, registry, file operations, strings, DLLs. Using these features, Ashraf *et al.* (2019) managed to obtain an accuracy of 92% for a support vector machine and 91% for a random forest using the dynamic information. The work by Fernando, Komninos, and Chen (2020) explored many ML algorithms that produced accuracies of between 33.1% and 96.4%, with the lowest being an SVM and the highest being a multi-layer perceptron with 10 hidden layers.

Findings in Arabo *et al.* (2017) explored malware detection using memory analysis of API calls, DLLs, process handles, network, code injection, and privilege as features in their dataset. The results obtained ranged from 95.94% to 98.50%, with the lowest being a random forest and the highest an SVM. However, the results for the random forest had a recall of 87.68%, which means it did not perform well in prediction. The SVM classifier on the other hand performed well and gave a low false-positive rate and recall. The study considered weighted features and had a large dataset of 3468 records. Similarly, Grégio *et al*. (2013) explored memory writes for malware classification and code-reuse identification. Clustering techniques were explored with a precision value of 84.3% using pre- and inter-clustering. To the best of the Author's knowledge, no research has been done using process memory specifically, rather process behaviour as seen in the work by Arabo *et al.* (2017). Their work involved collecting computer metrics such as the CPU, RAM, and disk usage, as well as the API calls and files open. The results of Arabo *et al.* (2017) were extremely poor and ranged from 52.85% for a neural network to 75.01% for a random forest. Existing literature (Dang *et al.*, 2021; Molina *et al.*, 2021; Rughani, 2017; Shah & Issac, 2018) shows there has been great reliance on the following ML algorithms: neural network; k-nearest neighbours; decision tree; random forest; Naïve Bayes; SVM; and boosted learners.

Exploring process memory specifically has been overlooked in the research space, since more data-rich features were available. However, these data-rich features often pose a degree of bias or allow attackers to easily bypass detection methods through deterrence mechanisms and obfuscation (Bashari Rad, Masrom & Ibrahim, 2012; Grégio *et al.*, 2013). Process memory can help identify the inner workings of any executable and with the power of machine learning, this can become simpler thereby minimizing manual analysis of large amounts of data. The current research paper aims to leverage process memory to distinguish ransomware more accurately from benignware. It also aims towards active detection without the need of utilizing a sandboxed environment.

## 3. Ransomware Detection using Process Memory

The proposed method of ransomware detection involved exploring the different memory regions that occur in process memory. The overall process employed in this research is modelled and shown in Figure 1. The process model starts with capturing the process's memory using the Cuckoo Sandbox (Guarnieri, 2014). Thereafter, Cuckoo runs a default reporting module that analyses the collected memory dump and produces a raw report with each memory region from process memory along with metadata. The access privileges that are occurring in each memory region are captured from each region. These privileges include Read (**r**), Read/Write (**rw**), Read/Execute (**rx**), Read/Write/Copy (**rwc**), Read/Write/Execute (**rwx**) and Read/Write/Execute/Copy (**rwxc**). The privileges determine what operations and what types of data are stored in memory to help find patterns in subroutines that can be attributed to specific operations. For example, if a segment within the memory region has an access privilege of **rx** and the type is **code**, this could be a subroutine involved with reading a file and then performing some computation on it. However, the privilege levels include several aspects, and one of the most interesting and rarely seen privileges is **rwxc**. Using this highest privilege level could mean that the executable can persist, extract, and execute code in memory. On the other hand, Read/Write/Execute (**rwx**) could indicate a dynamic subroutine-like encryption cipher. Process memory consists of many segments within a memory region and therefore the approach adopted in this research is a summation of each segment's privileges per memory region. Once the data has been collected, the next phase is to conduct the Machine Learning (ML) exploratory phase, which involves exploring various ML techniques for classification. The techniques this study explored are classifiers such as decision tree, random forest, gradient boost, XGBoost, tree ensemble, neural network, Naïve Bayes, and support vector machine (Fernando *et al.*, 2020). These algorithms were tested to see which provides the best accuracy in detecting ransomware and produces the least number of false negatives.

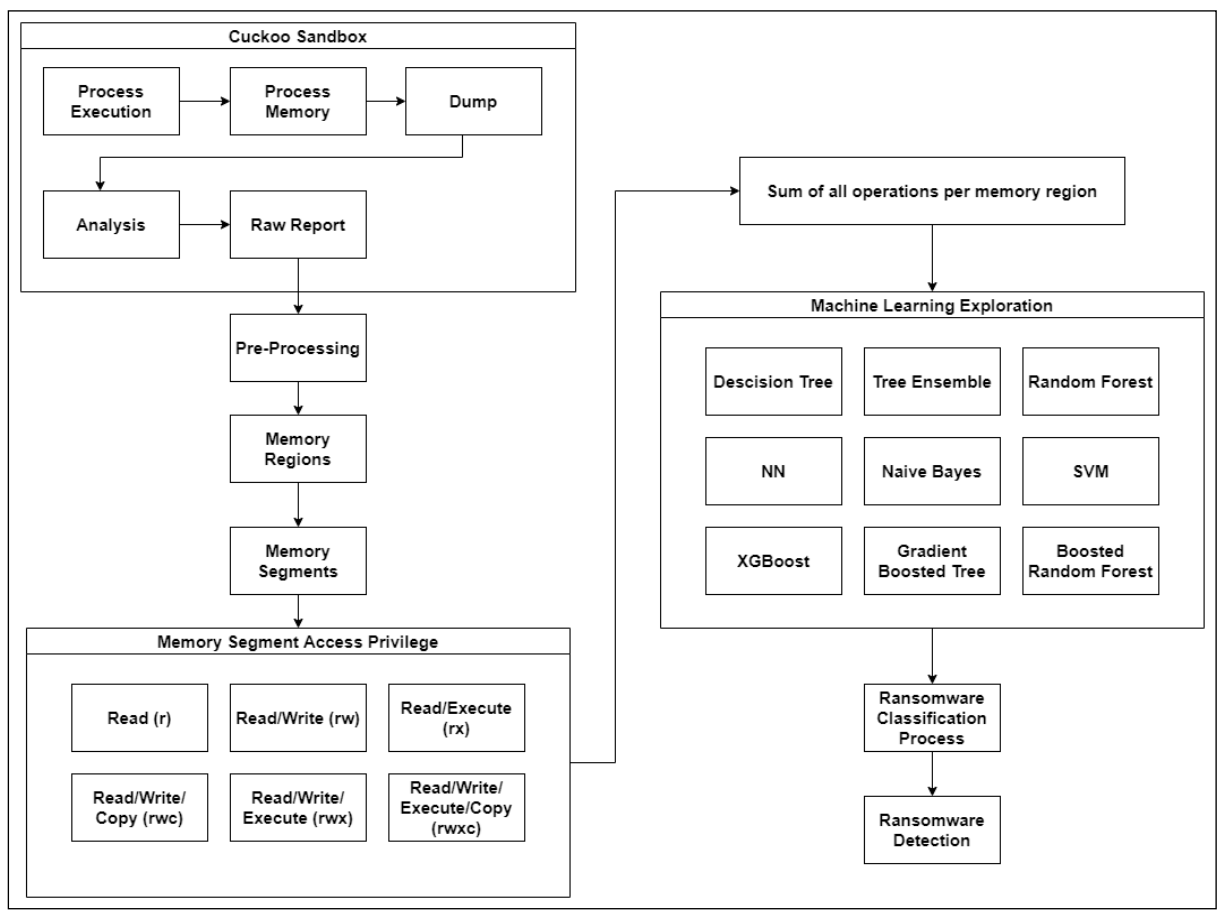

**Figure 1:** Ransomware Detection Process Model using Processor Memory

The subsections below discuss how the data was collected and pre-processed for the machine learning exploration phase. The results from the machine learning phase are also discussed.

### 3.1 Dataset

The most important aspect of machine learning is the feature engineering. Having enough information that represents all classes is extremely important to ensure a fair result in the learning phase. For this purpose, two classes were chosen for detection, namely benign and malicious.

*3.1.1   Data Acquisition and Extraction Process*

Acquiring the data involved a major process, which had to be done manually as there are no existing datasets that comprise process memory. Benign samples were obtained from various sources, with most originating from Portable Apps (Haller, 2021). A total of 354 benign samples were collected across different categories of executables. The categories explored were accessibility, development, education, games, image software, music/video software, office, security, utilities, internet, encryption, and general-purpose. These benign samples were taken at random from each category.

Malicious samples were sourced mainly from Malware Bazaar (Abuse.ch, 2020) and theZoo (Yuval Nativ, 2015). A total of 117 ransomware samples were sourced from over 70 categories of ransomware families. A snippet of the breakdown of each benign sample and malicious ransomware category is given in Table 1. The full dataset can be found at https://github.com/ICFL-UP/RDPM.

**Table 1:** Summary of samples in each category

| Benign | | Malicious | | | | | |
|---|---|---|---|---|---|---|---|
| Category | Count | Category | Count | Category | Count | Category | Count |
| accessibility | 7 | #Other | 5 | Clay | 1 | DarkSide | 3 |
| development | 20 | 4444 | 1 | Clop | 3 | DeathHiddenTear | 1 |
| education | 11 | ArkeiStealer | 4 | CobaltStrike | 1 | Demonware | 3 |
| games | 20 | Avaddon | 3 | Conti | 3 | Dharma | 3 |
| graphics_pictures | 32 | BB | 1 | CryLock | 3 | DoejoCrypt | 3 |
| internet | 48 | Babuk | 4 | Crypt888 | 1 | Dridex | 1 |
| music_video | 33 | Balaclava | 4 | CryptoDarkRubix | 1 | EKANS | 1 |
| office | 29 | BlackKingdom | 2 | CryptoLocker | 1 | ESCAL | 3 |
| security | 17 | BlackRose | 1 | Cryptowall | 1 | Egregor | 3 |
| utilities | 137 | Cerber | 1 | Cuba | 1 | Encrpt3d | 1 |

Cuckoo Sandbox (Guarnieri, 2014) was used to generate the process memory for each sample. Each sample was analysed with the default Cuckoo settings and the process memory was dumped and analysed with Cuckoo reporting modules to extract the process memory. The report included the regions, segments as well as the access privileges and type of content stored, start and end address, state, offset, and size. A snippet of the data obtained from Cuckoo can be seen in Figure 2. The current study explored the protect field, which contains the access privilege. It was observed to be suitable, as the other attributes remained constant or were dependant on how the operating system assigned memory (start/end addresses, and the offset).

```
"procmemory": [
    {
        "regions": [
            {
                "protect": "rw",
                "end": "0x00020000",
                "addr": "0x00010000",
                "state": 4096,
                "offset": 24,
                "type": 262144,
                "size": 65536
            },
```

**Figure 2**: Process memory data from Cuckoo

*3.1.2   Data Pre-Processing and Metrics*

Each sample's memory region constituted a single record in the dataset. Some samples had multiple memory regions, while some only had one. Each memory region segment's access privileges from the Cuckoo reports were summed up for each privilege, thus resulting in the dataset as shown in Table 2. The dataset contained 476 benign records and 461 malicious records for a total dataset of 937 records and a total of 80 categories.

**Table 2:** Dataset sample

| R | RW | RX | RWC | RWX | RWXC | LABEL | CATEGORY |
|---|----|----|-----|-----|------|-------|----------|
| 367 | 307 | 117 | 84 | 70 | 0 | B | Utilities |
| 107 | 96 | 43 | 31 | 5 | 0 | B | Utilities |
| 62 | 59 | 18 | 21 | 49 | 0 | B | Utilities |
| 13 | 13 | 4 | 5 | 4 | 0 | M | Cryptowall |
| 124 | 94 | 54 | 32 | 6 | 0 | M | DarkSide |
| 239 | 289 | 79 | 104 | 98 | 22 | M | DeathHiddenTear |

### 3.2 Machine Learning Exploration Phase

Several classifiers (machine learning techniques) were explored with a training/testing ratio of 80/20 as the de facto standard (Ashraf *et al.*, 2019; Fernando *et al.*, 2020). Stratified sampling was used to ensure that the split was not biased or uneven in terms of the label distribution. Each algorithm was evaluated using the Knime Analytics Platform (Knime, 2021) on a Microsoft Windows machine. Metrics such as True Positive (TP), False Positive (FP), True Negative (TN), Recall, Precision, Sensitivity, Specificity, F-Measure, and Accuracy, were used to evaluate the result. The details of the explored algorithms are presented in the subsections that follow.

#### 3.2.1 Decision Tree (DT)

A Decision Tree is one of the most commonly used classification algorithms in the security domain (Berrueta, Morato, Magana, *et al.*, 2019; Molina *et al.*, 2021). Trees provide a rule-based approach to classification, and since security relies on rules, this is often the preferred approach (Adeyemi, Razak, Salleh, *et al.*, 2016; Ernsberger, Ikuesan, Venter, *et al.*, 2017; Mohlala, Ikuesan & Venter, 2018). The parameters used for the DT comprised the quality measure that integrates the Gini Index, with no pruning, and a minimum number of five records per node. An accuracy of 93.62% (see Table 3) was observed. The number of incorrect classifications was 6:6 for benign:malicious samples respectively. This provides a fair misclassification. The Cohen's Kappa score of 0.8723 implies that although the DT model performed relatively better in classification. However, it is inefficient for detecting ransomware as there would be a 6.38% chance it would go undetected.

**Table 3**: Decision Tree Results

| Label | TP | FP | TN | FN | Recall | Precision | Sensitivity | Specificity | F-measure | Accuracy |
|-------|----|----|----|----|--------|-----------|-------------|-------------|-----------|----------|
| B | 90 | 6 | 86 | 6 | 0.9375 | 0.9375 | 0.9375 | 0.9347 | 0.9375 | |
| M | 86 | 6 | 90 | 6 | 0.9347 | 0.9347 | 0.9347 | 0.9375 | 0.9347 | |
| Overall | | | | | | | | | | 0.9362 |

#### 3.2.2 Tree Ensemble (TE)

Ensemble learning is a technique that leverages concrete finite sets of models and evaluates them to find the best-performing model. Tree Ensemble is a model that builds multiple tree models and then evaluates each to find the best one. For the TE, the information gain ratio split criteria were utilized with the number of models set to 100 with square root attribute sampling. Increasing the number of models did not affect the accuracy, and a test was done with 10 models, which only affected the accuracy by 0.22%. Changing the split criteria and the sampling attributes had little to no effect on the accuracy of 95.74% (see Table 4). This is an improvement over the DT classifier, thereby providing better classification and a decrease in the confusion matrix of just two misclassifications for each label.

**Table 4.** Tree Ensemble Results

| Label | TP | FP | TN | FN | Recall | Precision | Sensitivity | Specificity | F-measure | Accuracy |
|-------|----|----|----|----|--------|-----------|-------------|-------------|-----------|----------|
| B | 90 | 2 | 90 | 6 | 0.9375 | 0.9783 | 0.9375 | 0.9783 | 0.9574 | |
| M | 90 | 6 | 90 | 2 | 0.9783 | 0.9375 | 0.9783 | 0.9375 | 0.9574 | |
| Overall | | | | | | | | | | 0.9574 |

#### 3.2.3 Random Forest (RF)

Unlike a Tree Ensemble (TE), a Random Forest builds multiple tree structures within one model building a forest of trees. This classification algorithm is very useful for a complex hierarchy of rules and is particularly used often in the security domain for classification. While the RF did perform well it is slightly less accurate than the TE. This can be due to the dataset not having too many patterns from the different occurrences of the data points, resulting in a rather small forest, achieving a similar result as the TE. An RF and TE are very similar from an algorithm point of view, therefore, having accuracy in the same range is expected. The accuracy statistics of the RF are shown in Table 5.

Table 5: Random Forest Results

| Label | TP | FP | TN | FN | Recall | Precision | Sensitivity | Specificity | F-measure | Accuracy |
|---|---|---|---|---|---|---|---|---|---|---|
| B | 89 | 2 | 90 | 7 | 0.9271 | 0.9780 | 0.9271 | 0.9783 | 0.9519 | |
| M | 90 | 7 | 89 | 2 | 0.9783 | 0.9278 | 0.9783 | 0.9271 | 0.9524 | |
| Overall | | | | | | | | | | 0.9521 |

### 3.2.4 Gradient-Boosted Tree (GBT)

A Gradient-Boosted Tree is an ensemble learning where the notion of combining previous models would increase prediction. Hence, the goal is to minimize the gradient of the prediction error. Based on the results in Table 8, the accuracy is 94.68% with a false-positive misclassification of only four records. In comparison, DT had 6, whereas TE and RF had two, which means that GBT does not perform better. This is also verified by the accuracy score is lower than the other classifiers. Generally, it is said that gradient boosting gives a higher performance than a random forest, due to the optimization technique of reducing the prediction error.

Table 6: Gradient-Boosted Tree Results

| Label | TP | FP | TN | FN | Recall | Precision | Sensitivity | Specificity | F-measure | Accuracy |
|---|---|---|---|---|---|---|---|---|---|---|
| B | 90 | 4 | 88 | 6 | 0.9375 | 0.9574 | 0.9375 | 0.9565 | 0.9474 | |
| M | 88 | 6 | 90 | 4 | 0.9565 | 0.9362 | 0.9565 | 0.9375 | 0.9462 | |
| Overall | | | | | | | | | | 0.9468 |

### 3.2.5 XGBoost

The XGBoost is a gradient boost algorithm with additional features that allow for better performance and accuracy. These features include a robust penalty function on the tree models, shrinking of leaf nodes, and the ability to deal with large data. While GBT did not perform well (see Table 7), the results of the XGBoost algorithm delivered better results, with a 96.28% accuracy and a relatively high precision and F-measure score further validating the accuracy results obtained. The XGBoost algorithm was configured with the softprob option to give the probability for each data point. A boost of 1000 rounds improved the accuracy.

Table 7: XGBoost Results

| Label | TP | FP | TN | FN | Recall | Precision | Sensitivity | Specificity | F-measure | Accuracy |
|---|---|---|---|---|---|---|---|---|---|---|
| B | 93 | 4 | 88 | 3 | 0.9688 | 0.9588 | 0.9688 | 0.9565 | 0.9637 | |
| M | 88 | 3 | 93 | 4 | 0.9565 | 0.9670 | 0.9565 | 0.9688 | 0.9617 | |
| Overall | | | | | | | | | | 0.9628 |

### 3.2.6 Naïve Bayes (NB)

The Naïve Bayes algorithm tends to perform well, which results in higher accuracy than other algorithms. This is because the NB algorithm treats features individually, and these features are not related to each other. However, in this case, it performed poorly, which can be attributed partly to the smaller feature space and the fact that the features are related to one another. A Cohen's Kappa score of 0.6288 was obtained. The relatively low output indicates that this classification is not reliable. An accuracy score of 81.38% (see Table 6) further confirms that the NB algorithm is not efficient in detecting ransomware.

Table 8: Naive Bayes Results

| Label | TP | FP | TN | FN | Recall | Precision | Sensitivity | Specificity | F-measure | Accuracy |
|---|---|---|---|---|---|---|---|---|---|---|
| B | 71 | 10 | 82 | 25 | 0.7396 | 0.8765 | 0.7396 | 0.8913 | 0.8023 | |
| M | 82 | 25 | 71 | 10 | 0.8913 | 0.7664 | 0.8913 | 0.7396 | 0.8241 | |
| Overall | | | | | | | | | | 0.8138 |

### 3.2.7 Support Vector Machine (SVM)

The Support Vector Machine algorithm was tested given that SVM works well with small datasets such as the one used in the current study. The SVM performs classification based on hyperplanes so that each label can be isolated from the other. However, the SVM algorithm did not perform well with this dataset, and it achieved an accuracy of 85.64% (see Table 10). A test was done using the configuration of a polynomial kernel where the power, bias, gamma, and penalty were set to 1, which resulted in an accuracy of 52.45%. The final configuration (see the results in Table 10) used a radial basis function kernel with a sigma of 0.1.

**Table 9**: Support Vector Machine Results

| Label | TP | FP | TN | FN | Recall | Precision | Sensitivity | Specificity | F-measure | Accuracy |
|---|---|---|---|---|---|---|---|---|---|---|
| B | 94 | 25 | 67 | 2 | 0.9792 | 0.7899 | 0.9792 | 0.7283 | 0.8744 | |
| M | 67 | 2 | 94 | 25 | 0.7283 | 0.9710 | 0.7283 | 0.9792 | 0.8323 | |
| Overall | | | | | | | | | | 0.8564 |

#### 3.2.8 Neural Network (NN)

To compare a simple supervised learning classification to deep learning algorithms, a Neural Network was also explored. Resilient Propagation Multi-Layer Perceptron (RProp MLP) as defined by Riedmiller and Braun (1993) was used instead of a normal NN, because RProp MLP adjusts the weights of the network based on the behaviour of the error function. Since the data could not be used in its raw form, further pre-processing was needed for the NN training and all the features were normalised to values between 0 and 1. Different configurations of the parameters were explored, and the best performing parameters were 10000 iterations, two hidden layers, and 15 hidden neurons per layer. This resulted in an accuracy of 93.62% (see Table 11), which is on par with the tree classifiers.

**Table 10:** Neural Network Results

| Label | TP | FP | TN | FN | Recall | Precision | Sensitivity | Specificity | F-measure | Accuracy |
|---|---|---|---|---|---|---|---|---|---|---|
| B | 120 | 8 | 144 | 10 | 0.9231 | 0.9375 | 0.9231 | 0.9474 | 0.9302 | |
| M | 144 | 10 | 120 | 8 | 0.9474 | 0.9351 | 0.9474 | 0.9231 | 0.9412 | |
| Overall | | | | | | | | | | 0.9362 |

### 4. Discussion

Overall, process memory gave a good measure for ransomware detection as most ML algorithms gave an accuracy of 93-96% (see Table 12). The Neural Network classifier also provided a relatively good performance. However, further exploration of other deep learning classifiers was discouraged in this study. A deep learning algorithm requires a relatively larger dataset, whereas the dataset used in this study was insufficient for typical deep learning algorithms to make good predictions. For instance, NN needs a larger dataset to generate a more accurate classification. Getting more data is a difficult process, as samples need to be manually run and a Cuckoo report can easily extend over 1 GB for each ransomware sample. Expanding on the dataset would also need more benign samples to maintain the class balance, in which case benign samples do not generate much process memory information – as opposed to ransomware. Machine learning algorithms suggested in (Berrueta *et al.*, 2019; Molina *et al.*, 2021; Rughani, 2017; Vinayakumar, Soman, Velan, *et al.*, 2017) obtained accuracy scores that ranged between 88-98%, which are good results. In comparison, the proposed research achieved the highest accuracy of 96.28% using the XGBoost algorithm. The limitation of the current research is that the dataset did not contain significantly larger training data. This can be attributed to the lack of resources and the absence of timeous acquisition, processing, and extraction of the process memory. Furthermore, the region of memory considered in this study is relatively measurable and logically feasible to provide a measure of dissimilarity between benignware and ransomware. The exploration of these memory-rich features presents a novel approach to ransomware investigation, consequently, improving ransomware mitigation when used in digital forensic readiness.

**Table 11:** Summary of Accuracy Statistics

| TECHNIQUE | DT | TE | RF | NB | XG | GB | SVM | NN |
|---|---|---|---|---|---|---|---|---|
| Accuracy (%) | 93.62 | 95.74 | 95.21 | 81.38 | **96.28** | 94.68 | 85.64 | 93.62 |

To further evaluate the degree of reliability, the area under the Receiver Operating Characteristic (ROC) curve – AUC – was explored. The ROC measures binary classifiers on their discriminative threshold based on the ratio of a true positive rate to a false positive rate. The AUC provides a basis for an unbiased evaluation of the performance of a classifier relative to the baseline (random) classifier. The AUC typically ranges from 0 to 1, with 0 being the worst classifier performance. Figure 3 shows the ROC curve for the classifiers explored in this study, with the AUC for each classifier shown in parenthesis in the legend. Looking at Figure 3, the worst-performing model in terms of the AUC output is the NN. Overall, the best performing model in terms of accuracy and AUC is XGBoost (XG), with an AUC of 0.992 and an accuracy of 96.28%. Similarly, the performance of RF and TE respectively rivals XG. This echoes the observed accuracy in Tables 4 and 5, where both classifiers demonstrated similar performance.

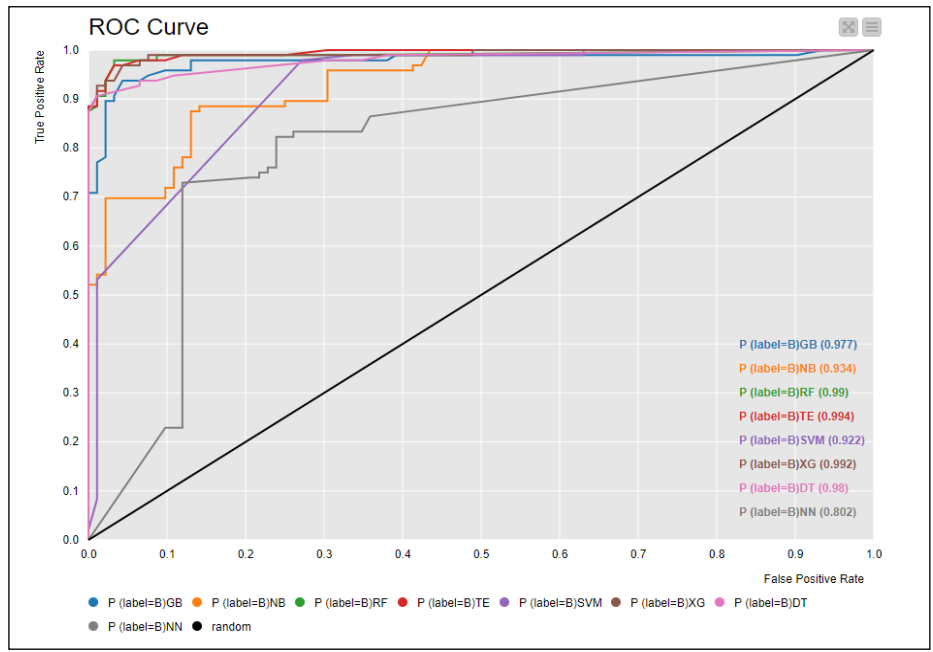

**Figure 3:** Area Under the ROC Curve of the Explored Classifiers

## 5. Conclusion and Future Work

In this paper, process memory was utilized for ransomware detection with the dataset comprising 937 records from 117 ransomware and 354 benign samples. The study observed, by leveraging several machine learning algorithms, that process memory does provide very meaningful information to assist with ransomware detection, and accuracy scores of between 93.62 and 96.28% were obtained, with the best performing model being XGBoost. Since static analysis often does not pick up ransomware attacks and behavioural analysis can be time-consuming, monitoring process memory is the most effective solution to actively detect ransomware attacks. Once process memory has been allocated, the access levels of the memory can provide meaningful information about the inner workings of the process. Data within the memory regions were not explored, as attackers would then make use of encryption to disguise their activity. Furthermore, the permissions of the access level cannot be hidden or encrypted, as this is done on an OS level. However, exploring the access level of a memory region is not sufficient for ransomware detection, therefore, future work would involve looking at the patterns in which these access levels occur for a more accurate and distinguishable mechanism for detection. Exploring the type of memory segments for each privilege can further provide more useful information to be used in prediction models. The paper in hand contributes to the security community by finding more novel ways for ransomware detection through automated behavioural analysis.